\documentclass[traditabstract]{aa}
\bibpunct{(}{)}{;}{a}{}{,} 
\usepackage{graphicx}
\usepackage[varg]{txfonts}
\usepackage{color}

\begin{document}

\title{Polarization of changing-look quasars\thanks{Based on observations made with the William Herschel telescope operated on the island of La Palma by the Isaac Newton Group of Telescopes in the Spanish Observatorio del Roque de los Muchachos of the Instituto de Astrofísica de Canarias and observations made with ESO Very Large Telescope at the Paranal Observatory under program ID~101.B-0209.}}
\author{D. Hutsem\'ekers\inst{1,}\thanks{Senior Research Associate F.R.S.-FNRS},
        B. Ag{\'\i}s Gonz\'alez\inst{1,2},
        F. Marin\inst{3},
        D. Sluse\inst{1},
        C. Ramos Almeida\inst{4,5},
        J.-A. Acosta Pulido\inst{4,5}
        }
\institute{
    Institut d'Astrophysique et de G\'eophysique,
    Universit\'e de Li\`ege, All\'ee du 6 Ao\^ut 19c,
    4000 Li\`ege, Belgium
    \and 
    Instituut voor Sterrenkunde, KU Leuven, Celestijnenlaan 200D, Bus 2401, 3001, Leuven, Belgium
    \and
    Universit\'e de Strasbourg, CNRS, Observatoire Astronomique de Strasbourg, UMR 7550, F-67000 Strasbourg, France
    \and
    Instituto de Astrofisica de Canarias, Calle Via Lactea, s/n, E-38205 La Laguna, Tenerife, Spain
    \and
    Departamento de Astrofisica, Universidad de La Laguna, E-38205 La Laguna, Tenerife, Spain
    }
\date{Received ; accepted: }
\titlerunning{Polarization of changing-look quasars} 
\authorrunning{D. Hutsem\'ekers et al.}
\abstract{If the disappearance of the broad emission lines observed in changing-look quasars originates from the obscuration of the quasar core by dusty clouds moving in the torus, high linear optical polarization would be expected in those objects. We then measured the rest-frame UV-blue linear polarization of a sample of 13 changing-look quasars, 7 of them being in a type 1.9-2 state. For all quasars but one the polarization degree is lower than 1\%. This suggests that the disappearance of the broad emission lines cannot be attributed to dust obscuration, and supports the scenario in which changes of look are caused by a change in  the rate of accretion onto the supermassive black hole. Such low polarization degrees also indicate that these quasars are seen under inclinations close to the system axis. One type~1.9-2 quasar  in our sample shows a high polarization degree of 6.8\%. While this polarization could be ascribed to obscuration by a moving dusty cloud, we argue that this is unlikely given the very long time needed for a cloud from the torus to eclipse the broad emission line region of that object. We propose that the high polarization is due to the echo of a past bright phase seen in polar-scattered light. This interpretation raises the possibility that broad emission lines observed in the polarized light of some type~2 active galactic nuclei can be echoes of past type~1 phases and not evidence of hidden broad emission line regions.
}
\maketitle
%
%
%

\section{Introduction}
\label{sec:intro}
Type~1 active galactic nuclei (AGNs) are characterized by both broad and narrow emission lines in their optical spectrum, while type~2 AGNs only show narrow emission lines. The ``unification model'' suggests that type~1 and type~2 AGNs are the same objects viewed under different inclinations (the angle between the system axis and the line of sight), the AGN core being obscured by an equatorial dusty torus in type~2 objects \citep{1993Antonucci}. A key argument in favor of the unification model was the discovery via spectropolarimetry of hidden broad line regions in type~2 AGNs, a polar scattering region providing us with a periscopic view of the obscured nucleus \citep{1985Antonucci,2005Zakamska,2016Ramos}.

Some rare AGNs have changed from type~1 to type~1.9-2 (type~1.9 AGNs still show faint broad H$\alpha$ emission), or vice versa \citep[e.g.,][]{1971Khachikian,1986Cohen,1989Goodrich}, some of them accomplishing a full cycle \citep[e.g.,][]{2016McElroy}. These changes are accompanied by a dimming (type~1 $\rightarrow$ 2) or brightening (type~2 $\rightarrow$ 1) of the continuum. Until recently, ``changing-look'' AGNs with appearing or disappearing broad emission lines (BELs) were essentially Seyferts, i.e., low-luminosity AGNs (e.g., Mrk1018, \citealt{2016Husemann}; Mrk590, \citealt{2014Denney}; NGC2617, \citealt{2014Shappee}). \citet{2015LaMassa} found the first high-luminosity AGN (quasar) changing from type~1 to type~1.9. Soon after, \citet{2016Runnoe}, \citet{2016Ruan}, and \citet{2016MacLeod} uncovered 12 other changing-look quasars in which BELs appeared or disappeared on timescales of years. More changing-look quasars have recently been identified by \citet{2017Gezari}, \citet{2018Yang}, \citet{2018Stern}, and \citet{2019MacLeod}. 

To explain these spectral changes, two main interpretations have been proposed. First, the changes are caused by modifications in the source of ionizing radiation, likely a variation in the rate of accretion onto the supermassive black hole (SMBH) \citep[e.g.,][]{1984Penston,2014Elitzur}. An intrinsic dimming of the continuum source reduces the number of photons available to ionize the gas, resulting in a net decrease in the emission line intensity. Second, the changes are caused by variable dust absorption along the line of sight to the continuum source and the broad-line region (BLR) due to the motion of individual dusty gas clouds in a clumpy torus \citep[e.g.,][]{1989Goodrich,1992Tran}. These two scenarios can be better discriminated in quasars than in Seyferts because the BLR is larger in quasars so that the lower limit on the crossing time of absorbing clouds increases up to decades \citep{2015LaMassa,2016MacLeod}. The size of the BLR is indeed proportional to $L^{0.53}$, $L$ being the 5100~\AA\ optical luminosity \citep{2013Bentz}. It follows that the variable dust absorption scenario is disfavored in quasars because the timescales of extinction variations due to dusty clouds moving in front of the BLR are factors 2-10 too long to explain the observed spectral changes \citep{2015LaMassa,2016MacLeod}. Moreover, \citet{2017Sheng} found large variations in the mid-infrared luminosity echoing the optical variations that occur during the change of look of ten AGNs, and argued that this behavior is inconsistent with the variable obscuration scenario due to the excessive amount of extinction needed and the too long obscuration timescale.

Quasar light is known to be linearly polarized at optical wavelengths, with significant differences between type~1 and type~2 objects. As discussed in \citet[][hereafter Paper~I]{2017Hutsemekers} a clear dichotomy is seen in the polarization degree: all type~1 quasars have low polarizations $p < 2\%$,  while all type~2 quasars have high polarizations $p > 2\%$. This dichotomy was established at rest-frame UV-blue wavelengths for quasars with redshifts $0.2 \leq z \leq 0.7$ and luminosities  $\log L_{\rm [OIII]}$(erg s$^{-1}) > 41.5$ that are comparable to the redshifts and luminosities of currently known changing-look quasars (Sect.~\ref{sec:sample}). It is also valid at higher redshifts \citep{2018Alexandroff}. On the other hand the dichotomy is less clear for lower luminosity, lower redshift Seyferts due to the stronger contamination by the host galaxy light that dilutes the polarization \citep{1994Kay,1983Yee,1994Kotilainen,2008Hamilton}. Following models initially developed for Seyferts in the framework of the unification model \citep[e.g.,][]{2004Smith,2011Batcheldor,2014Marin}, the polarization properties can be interpreted as scattering off two regions: an equatorial ring located inside the dusty torus at the origin of polarization parallel to the system axis, and a more extended polar scattering region at the origin of perpendicular polarization \citep[e.g.,][]{2005Zakamska,2008Borguet}. In type~1 quasars, seen at low inclinations, the continuum source and both scattering regions are seen by the observer resulting in a low polarization. In type~2 quasars, seen at high inclinations, the quasar core is hidden by the torus and only highly polarized polar-scattered light is seen. A change in polarization is thus expected from type~1 to type~2 depending on inclination, as actually observed in Seyfert galaxies \citep{2004Smith,2014Marin}, although polarization also depends on the torus half-opening angle, which most likely varies from object to object \citep{2011Ramos,2012Marin}.

\begin{figure}[t]
\centering
\resizebox{\hsize}{!}{\includegraphics*{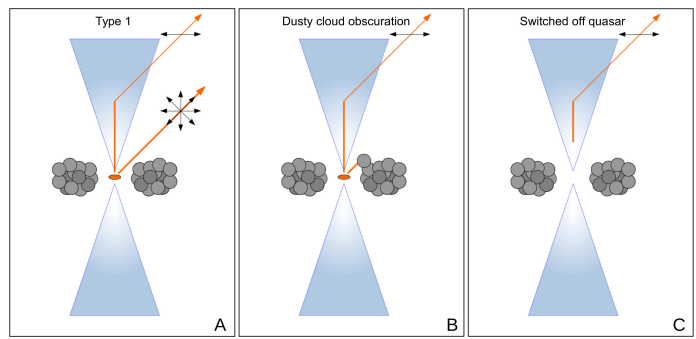}}
\caption{Cartoon representing the quasar core, torus clouds, and a polar scattering region. Light rays towards the observer are represented and their polarization illustrated by arrows. {\bf A:} The direct light from the quasar core (continuum source + BLR) reaches the observer and dilutes the light scattered off the polar regions. The quasar is in a type~1 state, and the polarization degree is low. {\bf B:} The direct light is blocked by a cloud intercepting the line of sight. The quasar is in a type~2 / obscured type~1 state, and the scattered light is no longer diluted. The polarization degree is high during the event.  {\bf C:} The quasar engine is switched off. Only the echo of a past bright state is observed in the scattered light. The polarization degree is high and slowly decreases with time. The quasar is in a true type~2 state.}
\label{fig:pola1}
\end{figure}

If the disappearance of broad emission lines in changing-look quasars is caused by torus clouds hiding the quasar core (i.e., the direct continuum source, the BLR, and the equatorial scattering region; see Fig.~\ref{fig:pola1} A\&B), only light scattered by the polar regions will reach the observer, so that the high polarization degree measured in type~2 obscured quasars is expected in quasars where the broad emission lines have disappeared.

In Paper~I, we measured the rest-frame UV-blue polarization of the changing-look quasar J101152.98$+$544206.4 in which the broad emission lines disappeared between 2003 and 2015. The polarization degree was found compatible with null polarization suggesting that the observed change of look was not due to a change in obscuration in the torus. Our results, substantiated by the simulations of \citet{2017Marin}, supported the idea that the vanishing of the broad emission lines in  J101152.98$+$544206.4 was due to an intrinsic dimming of the ionizing continuum source. In the present paper we report the measurement of the polarization of 12 other changing-look quasars. 

\section{Observations and measurements}
\label{sec:obs}

\subsection{Sample}
\label{sec:sample}

Our sample is constituted of all changing-look quasars discovered by \citet{2015LaMassa}, \citet{2016Runnoe}, \citet{2016Ruan}, and \citet{2016MacLeod}, i.e., a total of 13 quasars. The redshifts range from $z$ = 0.198 to $z$ = 0.625. All of them are high-luminosity AGNs  with $\log L_{\rm [OIII]}$(erg s$^{-1}) > 41.2$ \citep{2016Ruan, 2016MacLeod}. All targets are at high galactic latitudes ($|b_{\rm gal}| > 35\degr$), which ensures little contamination by interstellar polarization.  Polarimetry of J101152.98$+$544206.4 was previously reported in Paper~I.  

\subsection{Polarimetry and photometry with the VLT}

Polarimetric observations of eight quasars were carried out on September 13, 2018, using the ESO Very Large Telescope (VLT) equipped with FORS2 mounted at the Cassegrain focus of Unit Telescope \#1 (Antu). Linear polarimetry was performed by inserting a Wollaston prism in the parallel beam.  This prism splits the incoming light rays into two orthogonally polarized beams. Each object in the field therefore has two orthogonally polarized images on the CCD detector, separated by 22$\arcsec$. To avoid image overlapping, multi-object spectroscopy (MOS) slits were used to create a mask of alternating transparent and opaque parallel strips whose widths correspond to the splitting.  The final CCD image consists of alternate orthogonally polarized strips of the sky, two of them containing the polarized images of the object itself.  Because the two orthogonally polarized images of the object are recorded simultaneously, the polarization measurements do not depend on variable atmospheric transparency or on seeing. In order to derive the normalized Stokes parameters $q$ and $u$, blocks of four frames were obtained with the half-wave plate (HWP) at four different position angles: 0$\degr$, 22.5$\degr$, 45$\degr$, and 67.5$\degr$. While two different orientations of the HWP are sufficient to measure the linear polarization, the two additional orientations allow us to remove most of the instrumental polarization \citep{1989diSerego, 1999Lamy}. When exposure blocks were repeated, the HWP was rotated at the position angles 90$\degr$, 112.5$\degr$, 135$\degr$, and 157.5$\degr$. The targets were positioned at the center of the field to avoid the significant off-axis instrumental polarization generated by the FORS2 optics \citep{2006Patat}. All observations were carried out through the $g$\_HIGH+115 filter ($\lambda_{c}$ = 4670 \AA , FWHM = 1603 \AA ).

Data reduction and polarization measurements were done as described in \citet{2018Hutsemekers}. Basically, the $q$ and $u$ Stokes parameters are computed from the ratios of the integrated intensities of the orthogonally polarized images of the object, measured through aperture photometry for the four different orientations of the HWP. Since the Stokes parameters are usually found to be stable when varying the aperture radius, we adopted an aperture diameter of $n \times [(2 \ln 2)^{-1/2}\, \rm{FWHM}],$ where FWHM is the seeing value and $n = 3$. The polarization degree $p = (q^2 + u^2)^{1/2}$ and the associated error $\sigma_p \simeq \sigma_q \simeq \sigma_u $ \citep[see][for polarization statistics]{1958Serkowski,1962Serkowski} are given in Table~\ref{tab:data} for the observed quasars, together with the debiased polarization degree $p_0$ and the polarization position angle $\theta$. The debiased value $p_{0}$ of the polarization degree is obtained using the \citet{1974Wardle} estimator, accounting for the fact that $p$ is a positive quantity and then biased at low signal-to-noise ratio. The polarization position angle $\theta$ is derived by solving the equations $q = p\cos 2\theta$ and $u = p \sin 2\theta$. Due to the chromatic dependence of the HWP zero-angle, an offset is needed to convert the polarization angle measured in the instrumental reference frame into the equatorial reference direction (north = 0\degr ; east = 90\degr ).  For the $g$\_HIGH filter, this zero-point angle offset is 3.1$\degr$  according to the FORS2 user manual. This angle offset was checked using a polarized standard star (BD$-12\degr 5133$ from \citealt{2007Fossati}) and found within 1$\degr$ of its nominal value. The uncertainty of the polarization position angle $\theta$ is estimated from the standard \citet{1962Serkowski} formula where the debiased value $p_{0}$ is conservatively used instead of $p$, that is $\sigma_{\theta} = 28.65\degr \, \sigma_p / p_{0}$ \citep[see also][]{1974Wardle}. The instrumental polarization was not measured during that run by observing unpolarized standard stars, but it is lower than  0.1\% according to the FORS2 quality control \citep[see also][]{2018Hutsemekers}.

The type~1 / type~2 spectral changes are associated with large ($|\Delta g| > 1$ mag) photometric changes \citep{2016MacLeod}, and the quasars in our sample could have changed their type once again since the last observations reported in \citet{2016Ruan} and \citet{2016MacLeod}.  We then obtained direct images, deeper than the acquisition images, to perform relative photometry using objects in the quasar fields with $g$ magnitudes from the Sloan Digital Sky Survey (SDSS; \citealt{2000York}, \citealt{2011Eisenstein}). Typically more than five objects are used per quasar field to establish an approximate photometric zero point. The measured quasar magnitudes are reported in Table~\ref{tab:data} with errors estimated from the dispersion of the magnitudes measured for the field objects.

\subsection{Polarimetry and photometry with the WHT}

We  obtained linear polarization data of four other quasars on January 13, 14, and 15, 2018, using the Intermediate dispersion Spectrograph and Imaging System (ISIS) mounted at the Cassegrain focus of the 4.2m William Herschel Telescope (WHT) at the Roque de los Muchachos Observatory. Observations were done through the blue arm and the Sloan Gunn $g$ filter (ING filter \#218; $\lambda_c$ = 4844 \AA , FWHM = 1280 \AA ), with ISIS in its imaging polarimetry mode. As with FORS2, polarimetry with ISIS is performed by using a beam splitter, specifically a calcite Savart plate producing orthogonally polarized images in 6.4\arcsec \   sky strips separated by 7.7\arcsec. Blocks of four exposures with the HWP rotated at 7.5\degr, 52.5\degr, 30.0\degr, and 75.0\degr\ were secured. The weather was variable and exposure blocks repeated over several nights for the faintest targets. Given the similar instrumental setup, the measurement of the Stokes parameters was done as with FORS2 using the procedures described in \citet{2018Hutsemekers}, with the difference that the integrated intensities of the orthogonally polarized images of the object were measured using aperture diameters of $n \times [(2 \ln 2)^{-1/2}\, \rm{FWHM}]$ with $n$ smaller than 3, especially when the seeing was higher than 1.5\arcsec - 2\arcsec\  because the polarized sky strips containing the target images are narrower with ISIS (6.4\arcsec\ versus 22\arcsec\ with FORS2). The multiple exposure blocks were combined to increase the signal-to-noise ratio. The polarized standard stars BD$+25\degr 727$ (= HD283812) and BD$+59\degr 389$ \citep{1990Turnshek} were observed to correct for the chromatic dependence of the HWP zero-angle. Unpolarized standard stars (HD14069, G191B2B, and GD319 from \citealt{1990Turnshek}) were observed to estimate the instrumental polarization, found equal to $p = 0.04 \pm 0.03 \%$ after combining all measurements. The polarization measurements of the observed quasars are listed in Table~\ref{tab:data}. 

We also obtained direct images with the Auxiliary-port CAMera (ACAM) that is mounted permanently at a folded-Cassegrain focus of the WHT. The Sloan Gunn $g$ filter (ING filter \#701; $\lambda_c$ = 4660 \AA , FWHM = 1351 \AA ) was used. Relative photometry was performed using at least five objects of known SDSS $g$ magnitude in the quasar fields. The measured quasar magnitudes and their uncertainties are given in Table~\ref{tab:data}.

\begin{table*}
\caption{Polarization of changing-look quasars}
\label{tab:data}
\centering
\begin{tabular}{lccccccccrl}
\hline\hline
Object & $z$ & Observation Date &     $p$  & $\sigma_p$ & $p_0$ &  $\theta$ & $\sigma_{\theta}$  &  $g$   & $\Delta g$ \ & Spectral Type \\
       &     & (yyyy-mm-dd)     &   (\%)   & (\%)       & (\%)  &  ($\degr$)& ($\degr$)         &  (mag)  & (mag)     &   \\
\hline \\
J101152.98$+$544206.4 & 0.246 & 2017-02-19       & 0.15 & 0.22 & 0.00 & -   & -   &  19.6$\pm$0.2  & $0.0$        &  \ \ \ \ 2 $\rightarrow$ 2   \\
J015957.62$+$003310.5 & 0.312 & 2018-01-13/14/15 & 0.47 & 0.27 & 0.41 & 158 & 16  &  20.5$\pm$0.2  & $+0.5$       &  \ \ \ \ 2 $\rightarrow$ 2   \\
J100220.17$+$450927.3 & 0.400 & 2018-01-13/15    & 0.61 & 0.17 & 0.59 &  68 & 8   &  19.4$\pm$0.1  & $-1.0$       &  \ \ \ \ 2 $\rightarrow$ 1?  \\
J102152.34$+$464515.7 & 0.204 & 2018-01-14       & 0.16 & 0.23 & 0.00 &  -   & -  &  19.4$\pm$0.1  & $0.0$        &  \ \ \ \ 2 $\rightarrow$ 2   \\
J132457.29$+$480241.2 & 0.272 & 2018-01-14       & 0.17 & 0.13 & 0.13 & 158 & 22  &  18.4$\pm$0.1  & $-1.5$       &  \ \ \ \ 2 $\rightarrow$ 1?  \\
J214613.31$+$000930.8 & 0.621 & 2018-09-13       & 0.28 & 0.43 & 0.00 &  -  & -   &  20.8$\pm$0.2  & $+0.5$       &  \ \ \ \ 1 $\rightarrow$ ?   \\
J225240.37$+$010958.7 & 0.534 & 2018-09-13       & 1.10 & 0.66 & 0.94 & 115 & 20  &  21.2$\pm$0.1  & $0.0$        &  \ \ \ \ 2 $\rightarrow$ 2   \\
J233317.38$-$002303.4 & 0.513 & 2018-09-13       & 0.59 & 0.51 & 0.38 & 138 & 39  &  21.2$\pm$0.1  & $+1.5$       &  \ \ \ \ 1 $\rightarrow$ 2?  \\
J233602.98$+$001728.7 & 0.243 & 2018-09-13       & 0.21 & 0.12 & 0.18 &  47 & 19  &  20.3$\pm$0.1  & $-1.0$       &  \ \ \ \ 2 $\rightarrow$ 1?  \\ 
J002311.06$+$003517.5 & 0.422 & 2018-09-13       & 0.31 & 0.09 & 0.30 & 131 & 9   &  18.8$\pm$0.1  & $0.0$        &  \ \ \ \ 2 $\rightarrow$ 2   \\ 
J012648.08$-$083948.0 & 0.198 & 2018-09-13       & 0.13 & 0.07 & 0.11 &  24 & 18  &  19.1$\pm$0.1  & $ ? $        &  \ \ \ \ 2 $\rightarrow$ ?   \\ 
J022556.07$+$003026.7 & 0.504 & 2018-09-13       & 0.48 & 0.13 & 0.46 & 147 & 8   &  20.7$\pm$0.1  & $0.0$        &  \ \ \ \ 1 $\rightarrow$ 1   \\ 
J022652.24$-$003916.5 & 0.625 & 2018-09-13       & 6.87 & 0.64 & 6.84 &  71 & 3   &  22.8$\pm$0.2  & $+1.0$       &  \ \ \ \ 2 $\rightarrow$ 2   \\ 
\hline
\end{tabular}
\end{table*}

\section{Results and discussion}
\label{sec:result}

\subsection{Variability}

The quasar magnitudes reported in Table~\ref{tab:data} are compared to the $g$-band light curves published by \citet{2016MacLeod,2019MacLeod}.  For the quasars J012648.08$-$083948.0 and J233602.98$+$001728.7 discovered by \citet{2016Ruan}, photometry from the SDSS and Pan-STARRS DR1 (PS1 ; \citealt{2016Chambers}, \citealt{2016Flewelling}) archives has been considered.

The value of $\Delta g$ given in Table~\ref{tab:data} is an estimate of the difference between the magnitude we measured for a given quasar and the magnitude evaluated at the epoch of its last recorded spectrum (essentially between 2010 and 2014). The  magnitude at the epoch of the last recorded spectrum was evaluated by interpolating or extrapolating available light curves. The value of $\Delta g$ is expressed in steps of 0.5 mag given the large uncertainties. Positive values indicate dimming. Five quasars in our sample show a large ($|\Delta g| \gtrsim 1$) variation between 2010/2014 and 2018.

In the last column of Table~\ref{tab:data}, the left number gives the spectral type from the last available spectroscopic observations \citep{2015LaMassa, 2016Runnoe, 2016Ruan, 2016MacLeod, 2019MacLeod}. For simplicity we use ``2'' for objects of type~1.9-2 or with BELs that disappeared, and ``1'' for objects of type~1 or with BELs that appeared. As shown by \citet{2016MacLeod,2019MacLeod}, bright states correspond to spectral type~1 and faint states to type~1.9-2.  In particular, recent spectroscopic observations of J002311.06$+$003517.5 and J225240.37$+$010958.7 confirm the disappearance of BELs in agreement with their current faint state. For those two objects, spectra obtained at three epochs indicate that cycles may be present \citep{2019MacLeod}. In the absence of spectroscopic observations simultaneous to our polarization measurements, we estimate the current spectral type from the photometry, given the correlation between brightness and the appearance or disappearance of BELs that was established for the quasars of our sample. The current spectral type is the rightmost number in the last column of Table~\ref{tab:data}. Six quasars previously classified as type 1.9-2 are as faint or fainter than in 2010/2014, and thus are almost certainly still in the same spectral state. J233317.38$-$002303.4, in a type~1 state in 2010, has decreased in brightness by $\sim$1.5 mag reaching the magnitude it had when it was in a type~1.9-2 state in 2001 \citep{2016MacLeod}. This quasar is thus most likely again in a type~1.9-2 state. Three other quasars that were in a type~1.9-2 state in 2010/2014 are now brighter by more than 1 magnitude, reaching the magnitude they had in 2001/2003 when they were classified type~1, and thus may again be in type~1 state. Spectroscopy would have been the ideal confirmation of the spectral type, but the lack of it has no impact on the results (see discussion in next section). In the following we focus the discussion on the seven quasars that are most likely in a type~1.9-2 state.

\subsection{Polarization properties}

Except J022652.24$-$003916.5 discussed below, all the quasars in our sample have low rest-frame UV-blue polarizations $p_0 < 1\%$, including six type~1.9-2 objects.  If the BEL disappearance is due to a change in obscuration in the torus, we would expect a net increase of the polarization degree up to the values measured in other type 2 quasars, i.e.,  $p > 2\%$  \citep[Paper~I; ][]{2017Marin}. The absence of such a high polarization in the type~1.9-2 quasars in our sample supports previous conclusions based on independent arguments (see Sect.~\ref{sec:intro}) namely that the spectral changes are not due to a change in obscuration, but rather to an intrinsic dimming of the ionizing continuum source caused by a decrease in the SMBH accretion rate. The small polarization degrees recorded for most quasars in our sample also suggest that these objects are seen at low inclination (i.e., close to face on) so that the scattering regions appear essentially symmetric, that is, at inclinations far from lines of sight crossing the torus. According to the simulations carried out by \citet{2017Marin}, polarization degrees lower than 1\% (respectively 2\%) originate from AGNs seen at inclinations smaller than 15\degr\ (respectively 25\degr). This implicitly assumes that changing-look quasars possess scattering regions similar to those found in other AGNs (Sect.~\ref{sec:intro}). We note that these results are not affected by the uncertainty of the current spectral types because all quasars but one have a low polarization degree irrespective of their spectral classification and the majority of them have not significantly brightened.

Interestingly, for nearly all quasars in our sample the observed changes are better attributed to a variation in the SMBH accretion rate. Instead,  changes of look observed in Seyferts are convincingly explained by either variation in the accretion rate (e.g., Mrk1018, \citealt{2016Husemann}; Mrk590, \citealt{2014Denney}; NGC2617, \citealt{2014Shappee}) or variable obscuration (e.g., NGC7603, NGC2622, \citealt{1989Goodrich}; Mrk993, \citealt{1992Tran}).

\subsection{Highly polarized quasar J022652.24$-$003916.5}

The quasar J022652.24$-$003916.5 ($z = 0.625$, hereafter J0226) is in a type 1.9-2 faint state. It is even fainter than previously reported by \citet{2016MacLeod}. J0226 is the only quasar in our sample to show high continuum polarization (at the rest-frame wavelength $\sim$ 2900 \AA ) in agreement with the polarization measured in bona fide type~2 quasars (see Paper~I). J0226 is fainter by about 2.5 magnitudes since its type~1 state in 2001. Dilution of its current polarization by a factor of 10 would result in a polarization degree $< 1 \%$ typical of type~1 quasars. Obscuration by a dusty cloud blocking the continuum source and the BLR could thus explain such a high polarization and the change of look. However, the time needed by a cloud from the torus moving on a Keplerian orbit to eclipse the BLR is around 80 years for J0226 \citep{2016MacLeod}, that is, more than one order of magnitude longer than the timescale of the observed BEL disappearance ($\sim$~6 years in the quasar rest-frame). This makes the variable obscuration scenario rather problematic as an explanation for the change of look observed in J0226\footnote{ J0226 is the quasar in the sample with the longest BLR crossing time.}.

Another interpretation of the high polarization observed in J0226 could be the time delay expected between an intrinsic dimming of the continuum seen in direct light and the dimming of the scattered continuum seen in polarized light when the quasar is seen at intermediate to high inclinations. While the light scattered by the smaller ($\lesssim$ 0.1 pc) equatorial ring suffers little delay in comparison to the changing-look timescale, the light scattered in polar regions extending over tens to hundreds of parsecs \citep{1995Capetti, 2002Kishimoto, 2005Zakamska} still reaches the observer decades to centuries after the direct light (Fig.~\ref{fig:pola1} C). In this case the polarized light contains the echo of a past bright phase, diluted by a much fainter direct light, then resulting in a high polarization degree. In such a scenario, we would expect the polarization degree to slowly decrease with time, unless another change of look occurs.

\begin{figure}[t]
\centering
\resizebox{\hsize}{!}{\includegraphics*{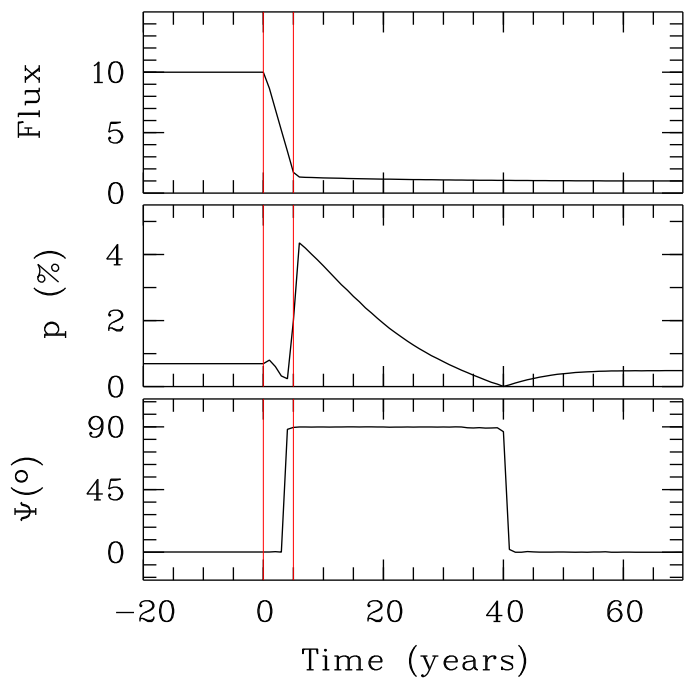}}
\caption{STOKES simulation of a type~1 quasar that suffers  dimming of the continuum source by a factor of 10 in 5 years (indicated by the red vertical lines). {\bf Top:} Total flux in arbitrary units seen by the observer as a function of time. {\bf Middle:} Evolution of the polarization degree ($p$ in percent). {\bf Bottom:} Evolution of the  polarization position angle ($\Psi$ in degrees).}
\label{fig:pola2}
\end{figure}

To establish this scenario on a more physical basis, we ran time-dependent simulations with the Monte Carlo radiative transfer code STOKES which models the polarization spectrum of an AGN including various sources and scattering regions \citep{2007Goosmann,2012Marin,2015Marin,2018Rojas,2018Marin}. In this particular case, we consider an unobscured type~1 AGN seen at an inclination of 45\degr. We assume a  standard model \citep{2017Marin}: an optically thick ($\tau_V$ > 50) dusty torus extending from 0.1 to 5 pc with a half-opening angle (as seen from the central SMBH) of 45\degr , an equatorial electron scattering region/BLR located just inside the torus with a half-opening angle of 20\degr, and a biconical, optically thin ($\tau_V \simeq$ 0.1), polar scattering region collimated by the torus half-opening angle and extending from 0.01 pc to 100 pc. The central continuum source emits unpolarized photons isotropically at the wavelength of 2900 \AA. The host galaxy is assumed to contribute 5\% of the quasar flux and is physically present in the simulation \citep[see][]{2018Marin}. The observed polarization is dominated by equatorial scattering and is typical of type~1 quasars (Fig.~\ref{fig:pola2}):  $p = 0.7\%$ parallel to the system symmetry axis ($\Psi$ = 0\degr). After some time, the source flux is assumed to drop by a factor of 10 within 5 years (red vertical lines in Fig.~\ref{fig:pola2}).  A complex polarization behavior is observed during that period that is due to the different time delays of the direct light, equatorially scattered light, and polar-scattered light.  When the source flux stabilizes in a faint state, the polarization has reached a high degree typical of type~2 quasars dominated by polar scattering  ($\Psi$ = 90\degr).  The polarization then slowly decreases over three decades, echoing the past bright phase. Since the bulk of the polarized flux actually comes from the inner ten parsecs of the polar scattering region \citep[see Fig.~6 of][]{2018Marin}, the timescale of the decay is on the order of decades. When the scattered light from the past bright phase finally vanishes, the system reaches a stable state dominated again by equatorial scattering with a polarization degree slightly smaller than at the beginning of the simulation due to the relative increase in the dilution by the host galaxy.

More generally, this scenario raises the interesting possibility that BELs observed in the polarized light of some type~2 AGNs can just be echoes of a past type~1 phase and not evidence of a hidden BLR (HBLR) surrounding an obscured type~1 core. Indeed, the BLR is small enough ($\lesssim$ 0.1 pc) to quickly react to a sudden dimming of the ionizing source so that BEL photons scattered in polar regions will be observed over decades in the polarized spectrum, while the AGN core is in fact in a faint, ``true'' type~2 state (i.e., a state where the absence of BELs is not due to the obscuration of a type~1 core). The presence of an HBLR in type~2 AGNs is usually considered as a pillar of the unification model (Sect.~\ref{sec:intro}).  Depending on the occurrence of the changing-look phenomenon in the AGN population, a more or less significant number of HBLR might need to be re-interpreted as echo BLRs, which would provide strong support to the existence of true type~2 AGNs.  This would also require adding another crucial parameter to the unification model, such as the accretion rate onto the SMBH.

\section{Conclusions}
\label{sec:end}

We report optical linear polarization measurements for a sample of 13 changing-look quasars. For all quasars but one, the rest-frame UV-blue polarization degree is lower than 1\%. Assuming that changing-look quasars have scattering regions similar to those found in other AGNs, low polarization degrees indicate that these objects are seen with a low inclination with respect to the line of sight. Low polarization degrees suggest that the change in quasar spectra from type~1 to type~1.9-2 cannot be attributed to dust obscuration, thus supporting the scenario in which changes of look are related to a change in accretion rate onto the SMBH.

One type~1.9-2 quasar in our sample shows a high polarization degree of 6.8\%. While this polarization could be ascribed to obscuration by a dusty cloud moving in the torus, we argue that this is unlikely given the very long time needed by a cloud from the torus to eclipse the BLR. We propose that, in this case, the high polarization is due to the echo of a past bright phase seen in polar-scattered light. This interpretation raises the possibility that BELs observed in the polarized light of some type~2 AGNs can be echoes of a past type~1 phase and not evidence of hidden BLRs.

\begin{acknowledgements}
We thank the referee for the constructive comments that helped to clarify the text.
DH and BAG thank Marie Hrudkova and Raine Karjalainen for their help during the WHT observations, and Elyar Sedaghati and Boris Haeussler for their help during the VLT observations. FM thanks the Centre national d'\'etudes spatiales (CNES) who funded his research through the post-doctoral grant ``Probing the geometry and physics of active galactic nuclei with ultraviolet and X-ray polarized radiative transfer.'' CRA acknowledges the Ram\'on y Cajal Program of the Spanish Ministry of Economy and Competitiveness through project RYC-2014-15779 and the Spanish Plan Nacional de Astronom\' ia y Astrofis\' ica under grant AYA2016-76682-C3-2-P.
This research has made use of the NASA/IPAC Extragalactic Database (NED), which is operated by the Jet Propulsion Laboratory, California Institute of Technology, under contract with the National Aeronautics and Space Administration.

Funding for SDSS-III has been provided by the Alfred P. Sloan Foundation, the Participating Institutions, the National Science Foundation, and the U.S. Department of Energy Office of Science. The SDSS-III web site is http://www.sdss3.org/. SDSS-III is managed by the Astrophysical Research Consortium for the Participating Institutions of the SDSS-III Collaboration including the University of Arizona, the Brazilian Participation Group, Brookhaven National Laboratory, Carnegie Mellon University, University of Florida, the French Participation Group, the German Participation Group, Harvard University, the Instituto de Astrofisica de Canarias, the Michigan State/Notre Dame/JINA Participation Group, Johns Hopkins University, Lawrence Berkeley National Laboratory, Max Planck Institute for Astrophysics, Max Planck Institute for Extraterrestrial Physics, New Mexico State University, New York University, The Ohio State University, Pennsylvania State University, University of Portsmouth, Princeton University, the Spanish Participation Group, University of Tokyo, University of Utah, Vanderbilt University, University of Virginia, University of Washington, and Yale University.

The Pan-STARRS1 Surveys (PS1) and the PS1 public science archive have been made possible through contributions by the Institute for Astronomy; the University of Hawaii; the Pan-STARRS Project Office; the Max-Planck Society and its participating institutes, the Max Planck Institute for Astronomy, Heidelberg, and the Max Planck Institute for Extraterrestrial Physics, Garching;  Johns Hopkins University; Durham University; the University of Edinburgh; the Queen's University Belfast; the Harvard-Smithsonian Center for Astrophysics; the Las Cumbres Observatory Global Telescope Network Incorporated; the National Central University of Taiwan; the Space Telescope Science Institute; the National Aeronautics and Space Administration under Grant No. NNX08AR22G issued through the Planetary Science Division of the NASA Science Mission Directorate; the National Science Foundation Grant No. AST-1238877; the University of Maryland; Eotvos Lorand University ; the Los Alamos National Laboratory; and the Gordon and Betty Moore Foundation.
  
\end{acknowledgements}

\bibliographystyle{aa}
\bibliography{references}

\end{document}